\documentclass[twocolumn]{aastex631}

\usepackage{amsmath}
 
 \usepackage {CJK}

\begin{document}
\begin{CJK*}{UTF8}{bsmi}
 
\title{Plasmoid-fed prominence formation (PF$^2$) during flux rope eruption} 

\author[0000-0003-2875-7366]{Xiaozhou Zhao (趙小舟)}
\affiliation{Centre for mathematical Plasma Astrophysics, Department of Mathematics, KU Leuven, \\
Celestijnenlaan 200B, 3001 Leuven, Belgium}

\author[0000-0003-3544-2733]{Rony Keppens}
\affiliation{Centre for mathematical Plasma Astrophysics, Department of Mathematics, KU Leuven, \\
Celestijnenlaan 200B, 3001 Leuven, Belgium}

\correspondingauthor{Xiaozhou Zhao}
\email{xiaozhou.zhao@kuleuven.be}

\begin{abstract}
We report a new, plasmoid-fed scenario for the formation of an eruptive prominence (PF$^2$), involving reconnection and condensation. We use grid-adaptive resistive two-and-a-half-dimensional magnetohydrodynamic (MHD) simulations in a chromosphere-to-corona setup to resolve this plasmoid-fed scenario. We study a pre-existing flux rope (FR) in the low corona that suddenly erupts due to catastrophe, which also drives a fast shock above the erupting FR.
A current sheet (CS) forms underneath the erupting FR, with chromospheric matter squeezed into it. The plasmoid instability occurs and multiple magnetic islands appear in the CS once the Lundquist number reaches $\sim 3.5\times 10^{4}$. The remnant chromospheric matter in the CS is then transferred to the FR by these newly formed magnetic islands. The dense and cool mass transported by the islands accumulates in the bottom of the FR, thereby forming a prominence during the eruption phase. More coronal plasma continuously condenses into the prominence due to the thermal instability as the FR rises. Due to the fine structure brought in by the PF$^2$ process, the model naturally forms filament threads, aligned above the polarity inversion line. Synthetic views at our resolution of $15 \mathrm{km}$ show many details that may be verified in future high-resolution observations.
\end{abstract}


\section{Introduction} \label{sec:intro}
Solar prominences (or filaments on the solar disk) are one of the most intriguing structures in  the solar atmosphere. 
Prominence plasma is roughly 100-fold cooler and 100-fold denser than the hot and tenuous solar coronal plasma, embedded in a typical magnetic field strength $3-30\,\mathrm{G}$~\citep{leroy1989observation}. The dense prominence plasma embedded in the coronal plasma is supported by magnetic fields.
The formation mechanisms of prominences are still puzzling.
There are various models accounting for prominence formation as reviewed by~\citet{mackay2010physics}, and the models include levitation of chromospheric plasma~\citep{Zhao2017,Zhao2019ApJ,Zhao2020ApJ}, evaporation-condensation~\citep{antiochos1999dynamic,Xia2011,Xia2014ApJ,Xia2016}, injection~\citep{an1988numerical,wang1999jetlike,guo2019formation}, or more mixed models such as levitation-condensation of purely coronal plasma~\citep{jenkins2021prominence},  as well as the three-dimensional prominence formation model by reconnection-condensation as proposed by~\citet{kaneko2017reconnection} and the prominence eruption model by~\citet{fan2018mhd}.

It is well acknowledged that solar flares and associated coronal mass ejections (CMEs) are different manifestations of a single physical process~\citep{Ko2003ApJ}, which is accompanied by prominence eruptions in many cases~\citep{martens1989circuit}.
The multi-physics evolution of flares is depicted in the observationally based CSHKP model for solar flares ~\citep{Carmichael1964,Sturrock1966,Hirayama1974,KoppPneuman1976}, which has recently been reproduced in self-consistent numerical simulations, where the magnetohydrodynamic (MHD) evolution interacted dynamically with advanced electron beam treatments~\citep{Ruan2020ApJ}.
Observations show that CMEs can exhibit a typical three-component morphology: an inner bright core of prominence material, a dark cavity, and a bright leading front~\citep{Illing1983,Illing1985,Webb2012}.
The progenitor of a CME is believed to involve a magnetic flux rope (FR).
The erupting FR evolves to the CME bubble and drives a bow shock in front of it, which may be identified as a bright leading front in observations~\citep{Lu2017ApJ835188L}. The fast-rising FR stretches the  magnetic field lines, and a current sheet (CS) is formed underneath. In our previous studies~\citep{Zhao2017,Zhao2019ApJ,Zhao2020ApJ}, we followed FR formation and eruption, accompanied by chromospheric plasma levitation, hence forming an erupting prominence, using two-and-a-half-dimensional MHD simulations. In those models, the initial linear force-free magnetic arcade was forced to form a FR by an imposed slow motion at the bottom boundary, converging towards the magnetic inversion line. This brings opposite-polarity magnetic flux to the inversion line, leading to the formation of an FR by magnetic reconnection. The FR erupts and lifts mass from the chromosphere to the corona, which forms an erupting prominence in a levitation scenario.

There are various triggering mechanisms for FR eruptions~\citep{Chen2011,keppens2019ideal}. For example, the FR eruption can be triggered by magnetic breakout~\citep{Antiochos1999}, shearing motion~\citep{Aly1990,Reeves2010ApJ72}, magnetic flux emergence~\citep{Chen2000}, or ideal MHD instabilities like the kink, torus or tilt instability models~\citep{Titov1999AA,Torok2005ApJ,Keppens2014ApJ77K}. 
The FR can also erupt due to loss of equilibrium through catastrophe, which is the triggering mechanism used in this paper, when it reaches its critical point~\citep{ForbesIsenberg1991}. The catastrophe model is analytically studied by~\citet{LinForbes2000} and numerically by~\citet{Mei2012,Ye2017AGUFMSH11B2436Y,Ye2020ApJ89764Y,Ye2021ApJ90945Y,Takahashi2017}. In the catastrophe model, there is already an FR before catastrophe, but then it suddenly forms a large-scale CS between the flare arcade and the CME, where magnetic reconnection takes place. This magnetic reconnection plays an essential role during the further eruption. 
However, classical two-dimensional steady reconnection models~\citep{dungey1953lxxvi,sweet1958production,Sweet1958,Parker1957,parker1963solar,Petschek1964} are either too slow to explain the flare time scale $\sim 100\,\mathrm{s}$, or not self-consistent in an MHD regime because anomalous resistivity is required in the diffusion region. 
Recent resistive MHD simulations~\citep{Barta2011,Ni2012ApJ75820N,Ni2018ApJ} demonstrate that a thinning Sweet-Parker CS can be unstable and fragment into multiple islands (or plasmoids), presenting fractal structure~\citep{ShibataTanuma2001}, once the Lundquist number exceeds a critical value of about $4\times10^4$~\citep{Bhattacharjee2009, Huang2010PhPl,Shen2011,Mei2012,Zhao2021arXiv210813508Z}. This yields a cascading process of the CS from large-scale structures to small-scale structures~\citep{Mei2012}.
The onset of the resulting plasmoid instability leads to a fast reconnection rate nearly independent of the Lundquist number. These plasmoids, carrying magnetic flux from the reconnection sites, play an important role for magnetic flux transport. Meanwhile, the plasmoids also transport mass and energy, and here we point out an as yet unanticipated role of plasmoids in the mass transport during a solar eruption: they can actually aid in forming a prominence.

In this paper, we report on a new scenario for prominence formation during eruption, namely a plasmoid-fed prominence formation (PF$^2$) model.
In this model, chromospheric matter is squeezed into the CS during its formation, and the newly formed magnetic islands in the CS carry this remnant chromospheric matter into the FR. Further coronal plasma condenses into the plasmoid-fed prominence as the FR rises.
The paper is organized as follows. Section~\ref{sec:model} describes the numerical setup. Section~\ref{rel} presents the numerical simulation results, where Section~\ref{glb} shows the global evolution of the FR eruption. Section~\ref{cspro} illustrates the prominence formation and eruption, which crucially involves the detailed CS evolution. Section~\ref{Forward} shows the synthetic images obtained in the forward modeling. Section~\ref{conclusion} summarizes the paper. 
 
\section{The MHD model} \label{sec:model} 
\subsection{The governing equations}

To investigate the FR eruption and the accompanied prominence formation, we solve the following set of resistive MHD equations:
\begin{equation}\label{eq:1}
\frac{\partial\rho}{\partial t}+\nabla\cdot \left(\rho\mathbf{u}\right)=0,
\end{equation}

\begin{equation}\label{eq:2}
\frac{\partial\left(\rho\mathbf{u}\right)}{\partial t}+\nabla\cdot \left[\rho\mathbf{u}\mathbf{u}+\left(p+\frac{\mathbf{B}^{2}}{8\pi}\right)\mathbf{I}-\frac{\mathbf{B}\mathbf{B}}{4\pi}\right]=  \rho\mathbf{g},
\end{equation}

\begin{equation} \label{eq:3}
\begin{split}
\frac{\partial \mathcal{H}}{\partial t}&+\nabla\cdot \left[\left(\mathcal{H}-\frac{\mathbf{B}^{2}}{8\pi}+p\right)\mathbf{u}+\frac{c\mathbf{E}\times\mathbf{B}}{4\pi}\right]\\&= \rho\mathbf{g}\cdot\mathbf{u} + \triangledown\cdot (\boldsymbol{\kappa}\cdot\triangledown T)-Q+h,	
\end{split}
\end{equation}

\begin{equation}\label{eq:4}
\frac{\partial\mathbf{B}}{\partial t}+\nabla\cdot \left(\mathbf{u}\mathbf{B}-\mathbf{B}\mathbf{u}  \right)= -c\nabla\times(\eta\mathbf{J}),
\end{equation}
where $\mathcal{H}=[\rho\varepsilon+(1/2)\rho\mathbf{u}^{2}+\mathbf{B}^{2}/8\pi]$ is the total energy density, $\varepsilon=p/(\gamma\rho-\rho)$ is the internal energy per unit mass, $\gamma=5/3$ is the adiabatic index, $\mathbf{E}=[\eta\mathbf{J}-(\mathbf{u}\times\mathbf{B})/c]$ is the electric field, $\eta$ is the resistivity, $\mathbf{J}=(c/4\pi)\nabla\times\mathbf{B}$ is the electric current density, $Q$ is the radiative loss rate, $h$ is the background heating rate, $\mathbf{g}$ is the gravitational acceleration, and $c$ is the light speed in vacuum. The set of MHD equations is closed by the equation of state
\begin{equation}
	p=\frac{\rho k_{\mathrm{B}}T}{\bar{\mu} m_{\mathrm{H}}},
\end{equation}
where $m_{\mathrm{H}}$ is the hydrogen atom mass and the mean molecular weight $\bar{\mu} = 1.4/2.3$ for the fully ionized plasma with a $10:1$ abundance of hydrogen and helium. The gravitational acceleration $\mathbf{g}$, the radiative cooling term $Q$, and the thermal conduction are treated in the same way as~\citet{Zhao2017,Zhao2019ApJ,Zhao2020ApJ}. The heating function $h$ is chosen to initially compensate the radiative loss exactly, i.e., $h(x,y)=Q(\rho_{I},T_{I})$, where $\rho_{I}(x,y)$ and $T_{I}(x,y)$ are the initial density and temperature distributions.

\subsection{Numerical setups}

The above equations are solved in two-and-a-half dimension, in a dimensionless form using the {\it MPI-AMRVAC} code~\citep{Nool2002,Porth2014ApJS,Xia2018ApJS,keppens2021mpi}.
To non-dimensionalize the equations, each variable is divided by its normalizing unit. The normalizing units are given in Table~\ref{Units}, and the units for derived variables are listed as well. The CGS-Gaussian units are used throughout the paper.  
The MHD equations are numerically solved with a finite-volume scheme setup combining the total variation diminishing Lax-Friedrich (TVDLF) scheme~\citep{yee1989class,Toth1996JCoPh} with a third-order Cada limiter~\citep{Cada2009JCoPh} for reconstruction, and a strong stability preserving Runge-Kutta four-step, 3rd-order time integration (SSPRK4), which is stable up to CFL of $\sim 2$~\citep{spiteri2002new}.

Our simulation is conducted in a two-dimensional Cartesian simulation box covering the region $[-25L_{0},25L_{0}]\times [ 0, 50L_{0}]$ in the {\it x-y} plane, where $L_{0}=10^{9}\,\mathrm{cm}$ is the typical length scale as listed in Table~\ref{Units}. The simulation box is divided into $256\times 256$ cells, which form the base computational grid.  Up to $8$ adaptive mesh refinement (AMR) levels are used, and thus the finest grid size reaches $0.0015L_{0}=15\,\mathrm{km}$.

\begin{table*}[htbp] 
	\begin{center}
\caption{Normalization Units.\label{Units}}
        {
\begin{tabular}{r r r r}
\hline\hline
Symbol & Quantity & Unit & Value\\
$x,y,z$ &Length & $L_{0}$ & $10^{9}\,\mathrm{cm}$\\

$T$ &Temperature & $T_{0}$ &  $1.0\times 10^{6} \,\mathrm{K}$ \\
$n$ & Number density & $n_{0}$ & $1.0\times 10^{9} \,\mathrm{ cm^{-3}}  $\\
$\rho$ & Mass density & $\rho_{0}=1.4n_{0}m_{\mathrm{H}}$ & $2.3417\times 10^{-15} \,\mathrm{g\cdot cm^{-3}}  $\\
$p$ &Pressure & $p_{0}=(\rho_{0} k_{B} T_{0})/(\mu_{w} m_{\mathrm{H}}) $ & $0.3175 \, \mathrm{Ba}$ \\
$\mathcal{H}$ & Energy density&  $p_{0}$ & $0.3175 \, \mathrm{erg\cdot cm^{-3}}$ \\
$\mathbf{B}$ &Magnetic induction& $B_{0}=\sqrt{4\pi p_{0}}$ & $1.9976\,\mathrm{ Gauss}$ \\
$\mathbf{u}$ &Velocity & $v_{0}=B_{0}/\sqrt{4\pi \rho_{0}} $ & $1.1645 \times 10^{7} \,\mathrm{cm\cdot s^{-1}}$\\

$t$ &Time&  $t_{0}=L_{0}/v_{0}$ &$85.8746 \,\mathrm{s}$ \\

$\eta$ & Resistivity&  $\eta_{0}=(4\pi L_{0}^{2})/(c^{2}t_{0})$ & $1.6282\times 10^{-4} \,\mathrm{s}$ \\

$\mathbf{E}$ & Electric field&  $(B_{0}L_{0})/(t_{0}c) $ & $7.759\times 10^{-4}\, \mathrm{statvolt/cm}$ \\

$\mathbf{J}$ & Current density&  $(B_{0}c)/(4\pi L_{0}) $ & $4.7689\,\mathrm{statamp/cm^{2}}$ \\
\hline
\end{tabular}}
	\end{center}
\end{table*}

 \begin{figure*}[htbp] 
	\epsscale{1.11}
    \plotone{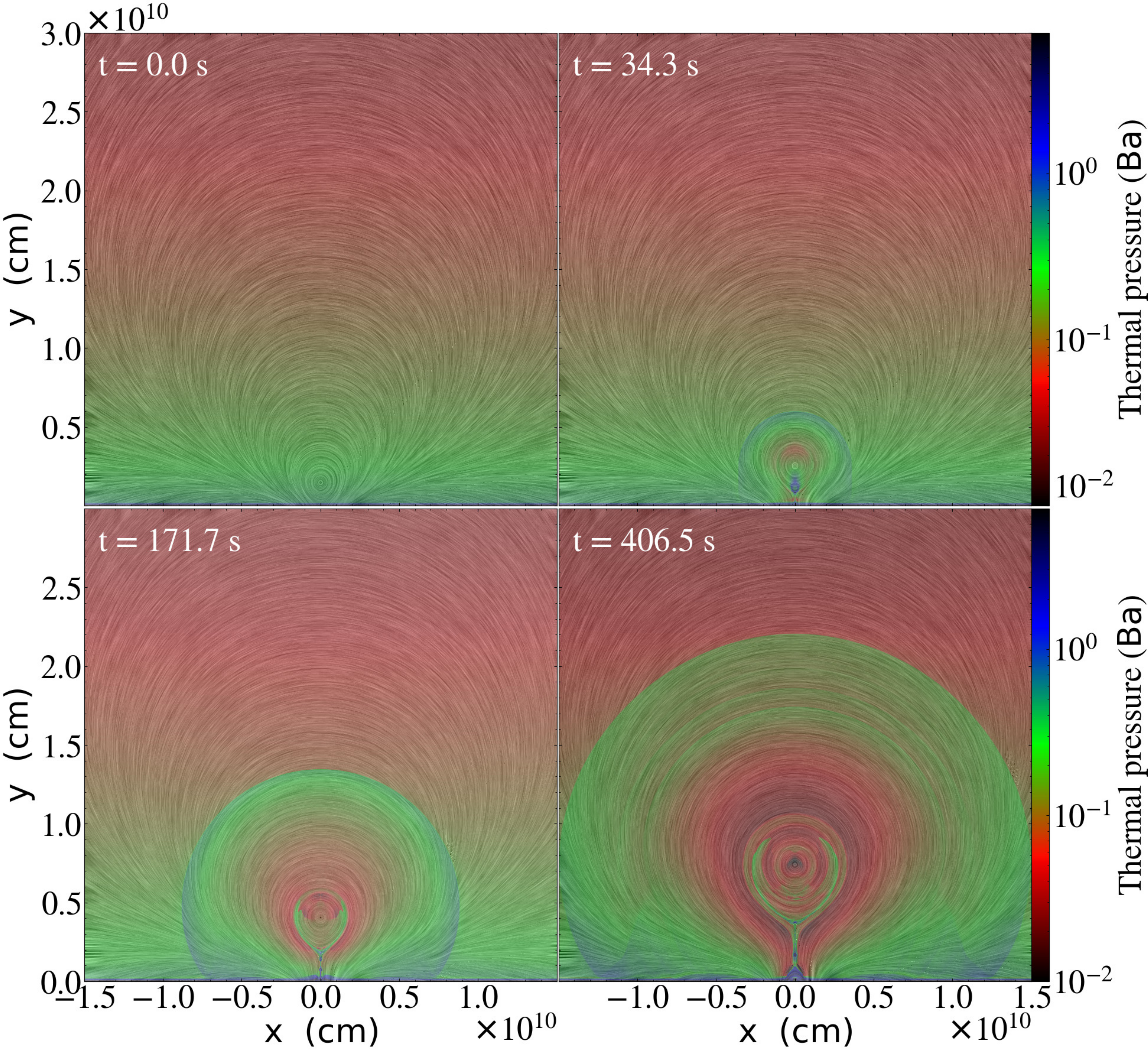}
    \caption{Thermal pressure evolution with magnetic field lines overlaid (in grey).\label{thp}}
\end{figure*}

\begin{figure*}[hbt!]
	\epsscale{1.2}
    \plotone{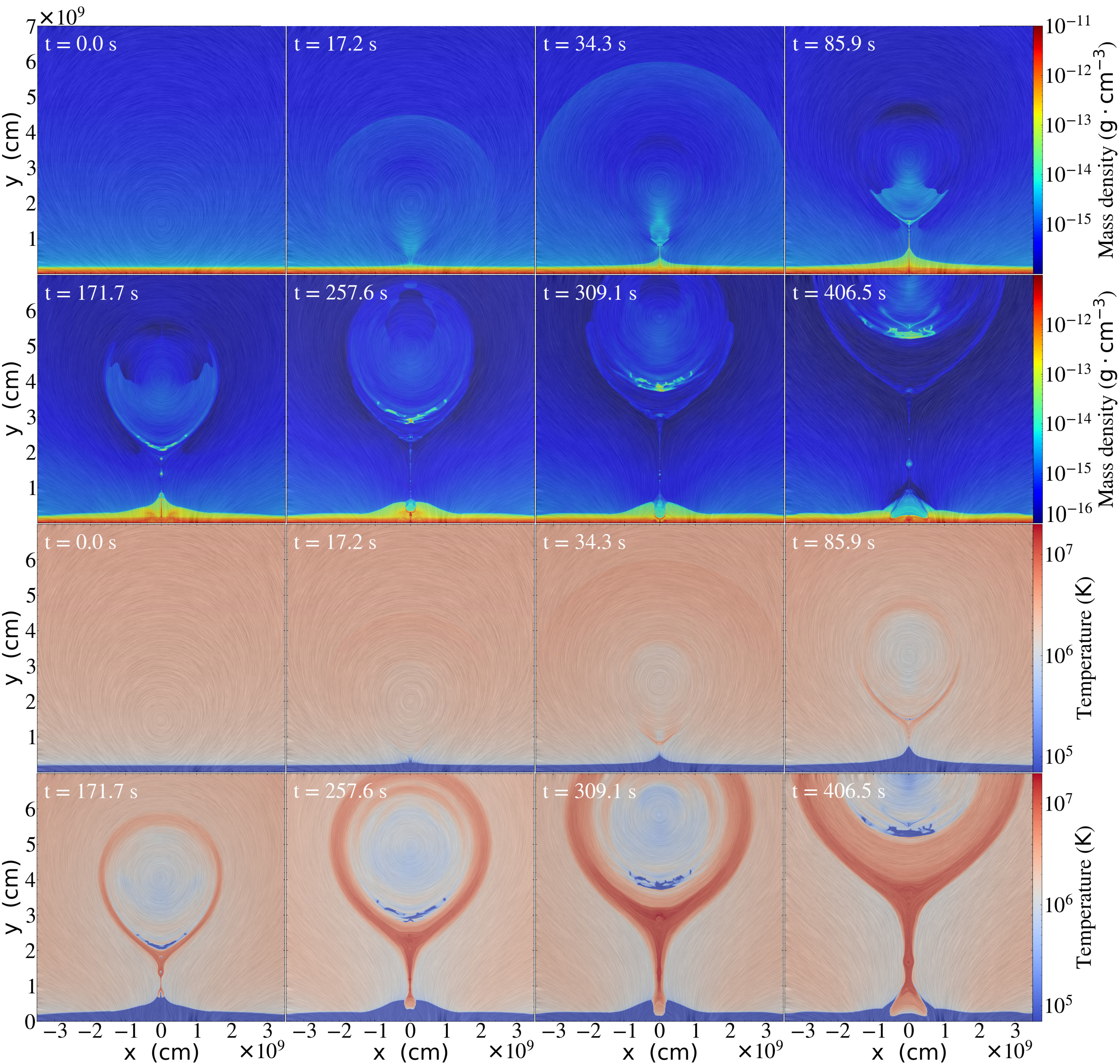}
    \caption{The evolution of mass density (first and second rows) and temperature (third and fourth rows) with magnetic field lines overlaid (in grey). \label{csE}}
\end{figure*}

\begin{figure*}[ht!] 
	\epsscale{1.2}
    \plotone{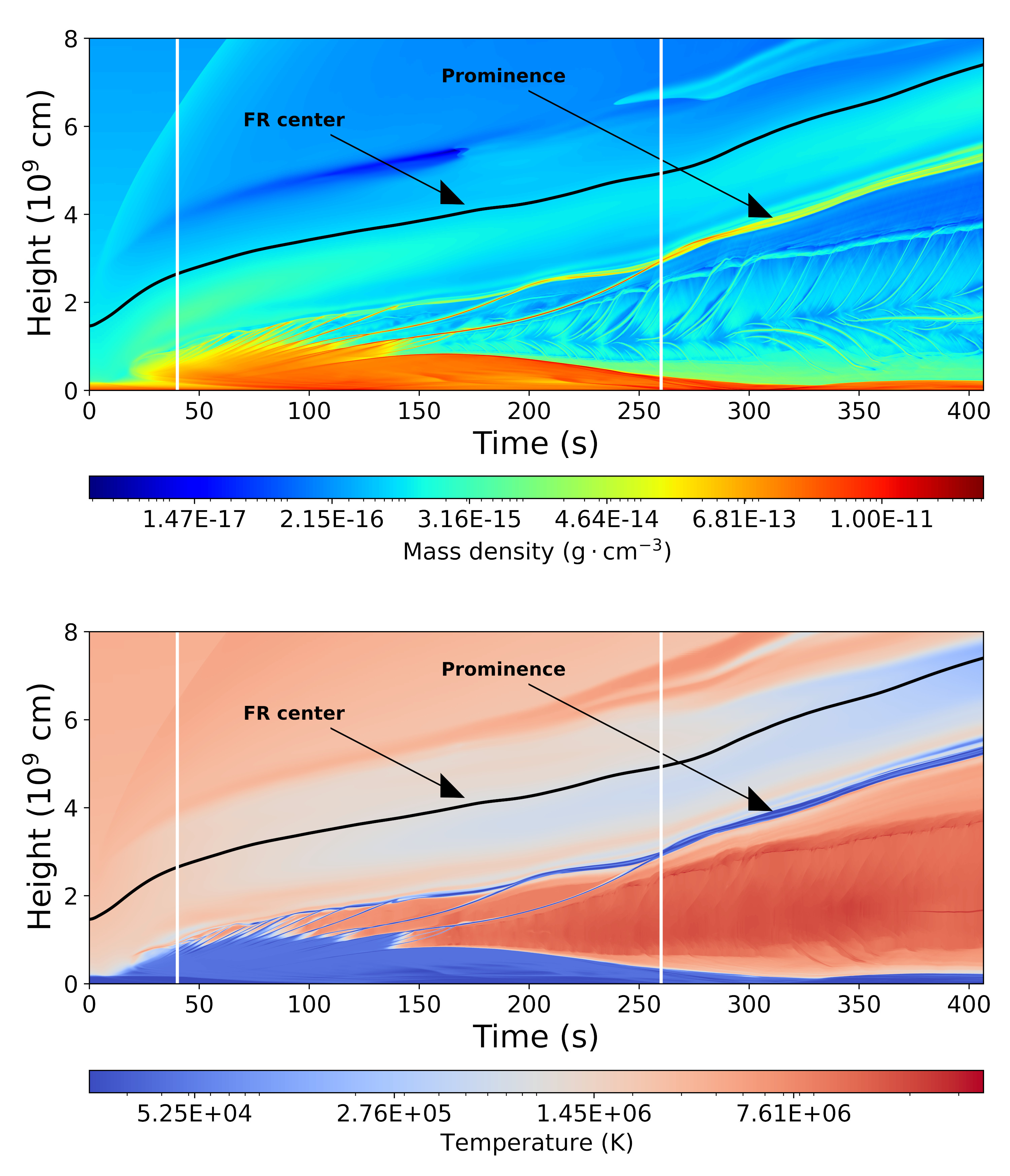}
    \caption{Time-distance diagrams of mass density (top) and temperature (bottom) along the line $x=0$ in the height range $y  \in  [0, 8L_{0}]$, where $L_{0}=10^{9}\,\mathrm{cm}$ is the typical length scale as listed in Table~\ref{Units}. The height of the FR center (black curve) and the prominence are indicated by arrows. The vertical solid lines indicate the instants $t=40\,\mathrm{s}$ and $t=260\,\mathrm{s}$, where the speed of the FR (center) changes.\label{hfr}}
\end{figure*}  
 
 \begin{figure*}[hbt!]
	\epsscale{1.2}
    \plotone{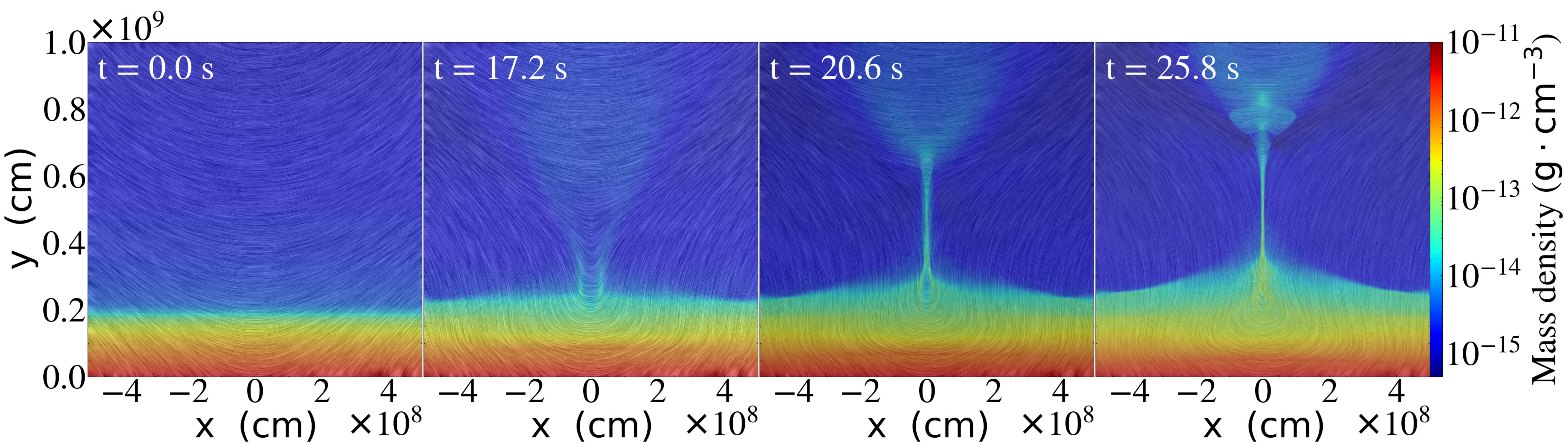}
    \caption{The evolution of mass density with magnetic field lines overlaid (in grey), illustrating the CS formation. \label{rev}}
\end{figure*}  
 
 \begin{figure*}[hbt!]
	\epsscale{1.2}
    \plotone{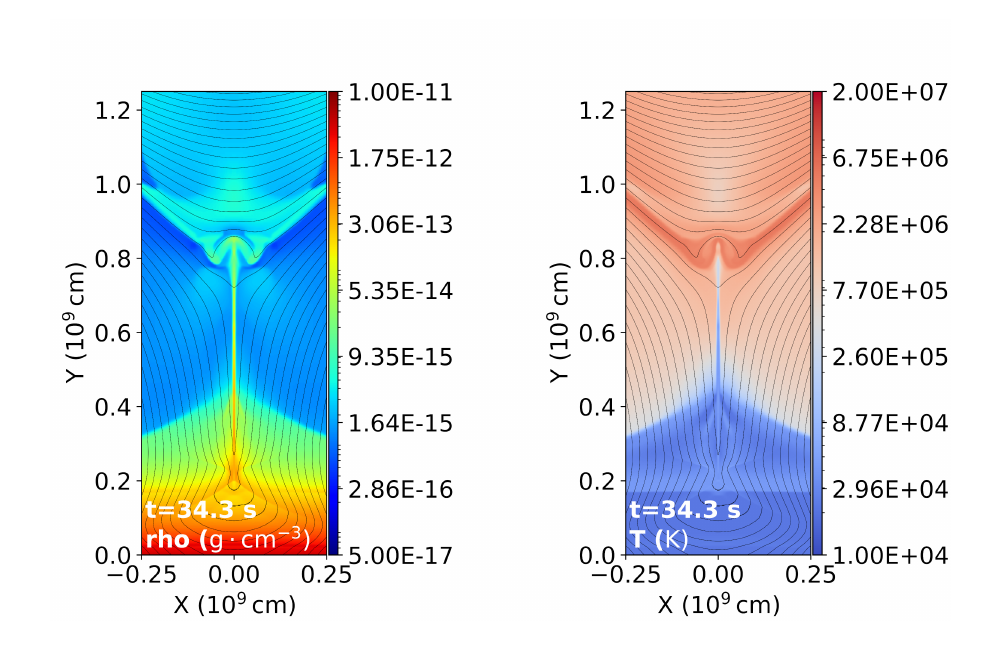}
    \caption{The zoomed view of the CS region at $t=34.3\,\mathrm{s}$ with magnetic field lines overlaid (black). \label{cszoom}}
\end{figure*}

\subsection{The initial and boundary conditions}

We adopt the same initial magnetic field configuration as in~\citet{Takahashi2017}, which is also provided in Appendix for convenience. We set the parameter $M_{q}$ as $0.8M_{c}$, with $M_c$ the critical value below which no FR equilibrium exists. The definitions of $M_{q}$ and $M_{c}$ are given in~\citet{Takahashi2017} as well as in Appendix. The parameters $A_{0}$, $R_0$, $\delta$ and $H$ in Appendix are set as $A_{0}=20\,\mathrm{Gauss}$, $R_0=1.2L_{0}$, $\delta=0.1L_{0}$ and $H=R_0$, respectively, where $L_{0}=10^{9}\,\mathrm{cm}$ is listed in Table~\ref{Units}. The initial temperature distribution is chosen in line with~\citet{Zhao2017,Zhao2019ApJ,Zhao2020ApJ}. The height of the chromosphere-corona transition region is set as $0.2L_{0}=2\,\mathrm{Mm}$ initially. The chromosphere with a temperature of ${10}^{4}\,{\rm{K}}$ is located below the transition region, which is a thin layer compared to the large size of the simulation domain. The coronal temperature profile is chosen as a vertically stratified profile with a constant vertical thermal conduction flux (i.e., $\kappa \partial T/\partial y=2\times {10}^{5}\,\mathrm{erg}\,{\mathrm{cm}}^{-2}\,{{\rm{s}}}^{-1}$, where $\kappa ={10}^{-6}{T}^{5/2}\,\mathrm{erg}\,{\mathrm{cm}}^{-1}\,{{\rm{s}}}^{-1}\,{{\rm{K}}}^{-3.5}$) above $0.2L_{0}=2\,\mathrm{Mm}$, following~\citet{Xia2012}. The initial density and pressure are then determined by assuming a hydrostatic atmosphere with a number density of ${10}^{14}\,{\mathrm{cm}}^{-3}$ at the bottom ghost cell boundary. 
 An FR with the same density and temperature of the coronal plasma pre-exists in the low corona in this setup, while, in contrast, the initial configuration is a linear force-free arcade in~\citet{Zhao2017,Zhao2019ApJ,Zhao2020ApJ}. The resistivity is set as a uniform and constant parameter $\eta=2\times10^{-5}\eta_{0}=3.2564\times10^{-9}\,\mathrm{s}$ throughout the simulation domain, which gives a global Lundquist number $S_{Lu}\simeq 5\times 10^{5}$ in the global length scale $\sim 10 L_{0}=100\,\mathrm{Mm}$. Here $\eta_{0}=1.6282\times 10^{-4} \,\mathrm{s}$ is also listed in Table~\ref{Units}.

The top and bottom boundaries of the simulation domain are handled mostly in the same manner as~\citet{Zhao2017,Zhao2019ApJ,Zhao2020ApJ} while open boundary conditions are applied to the left and right boundaries. In contrast to that earlier work, in this study, there is no driving velocity field, so the bottom boundary velocity is set to zero and is thus fixed.

 \section{Results}\label{rel}
 
 \subsection{Global evolution}\label{glb}

Figure~\ref{thp} presents the global evolution of the thermal pressure with magnetic field lines overlaid (in grey) in the region $[-15L_{0},15L_{0}]\times[0,30L_{0}]$ while Figure~\ref{csE} shows the density (first and second rows) and temperature (third and fourth rows) evolutions in a smaller region $[-3.5L_{0},3.5L_{0}]\times[0,7L_{0}]$. The chromosphere in Figure~\ref{thp} manifests as a thick line at the bottom because the height of the chromosphere-corona transition region $0.2L_{0}=2\,\mathrm{Mm}$ is much smaller than  $30L_{0}=300\,\mathrm{Mm}$, the size of the domain to illustrate. The chromosphere is clearly illustrated in Figure~\ref{csE} where the size of the domain shown is reduced. The parameter $M_{q}$ governs the force balance of the FR as discussed in~\citet{Takahashi2017} and in our Appendix. No equilibrium can be found in our setup since $M_{q}=0.8M_{c}<M_{c}$ and the FR is ejected upward immediately after the launch of the simulation due to catastrophe. Figure~\ref{hfr} shows the time-distance diagrams of mass density (top) and temperature (bottom) along the vertical line $x=0$ in the height range $[0,8L_{0}]$, respectively, where $L_{0}=10^{9}\,\mathrm{cm}$ is the typical length scale as listed in Table~\ref{Units}. The height of the FR center (the O-point) is also plotted over time in Figure~\ref{hfr}. The kinematic evolution of the FR is divided into three phases.  
The FR first goes through an initial acceleration due to the non-equilibrium initial setup until $t\sim 40\,\mathrm{s}$, which is indicated by the first vertical line in Figure~\ref{hfr}. Then the FR rises at a constant speed of $\sim 10^{7}\,\mathrm{cm\cdot s^{-1}}$ in the interval $40\,\mathrm{s}\leq t \leq 260\,\mathrm{s}$ (the region between the two vertical lines in Figure~\ref{hfr}). The speed of the FR rise then increases from $10^{7}\,\mathrm{cm\cdot s^{-1}}$ to $\sim 3\times 10^{7}\,\mathrm{cm\cdot s^{-1}}$ after $t\sim 260\,\mathrm{s}$ as indicated by the second vertical line in Figure~\ref{hfr}. The kinematic evolution of the FR is significantly different from what is described in~\citet{Zhao2017,Zhao2019ApJ,Zhao2020ApJ}, where the FR is formed and then undergoes a series of quasi-static equilibrium states in the initiation phase, followed by an impulsive acceleration. This difference is attributed to the different initial setups: the non-equilibrium initial setup in the current study imposes an initial acceleration to a pre-existing FR, while when an FR is formed due to converging motion, we get a quasi-static process where force balance is almost preserved in the CME initiation. In both cases though, a CS is formed underneath the FR. Below the reconnecting CS, flare loops are formed as a result of reconnection, which is consistent with the standard solar eruption model. As the reconnection continues, the flare loop system expands and its two foot-points separate. This flare foot-points separation was also reported in observations like~\citet{Qiu2002ApJ5651335Q} and the fully self-consistent MHD-beam flare model by~\citet{Ruan2020ApJ}.
The fast-rising FR drives the formation of a fast-mode shock straddling over it, as shown in Figure~\ref{thp} and \ref{csE}. These structures expand outward continuously as the FR rises.
\deleted{The fast shock front is also indicated in Figure~\ref{hfr}, which propagates at a speed of $4\times 10^{7}\,\mathrm{cm\cdot s^{-1}}$. The intersection lines between the front surface of the fast-mode shock and the chromosphere sweep the solar surface as the eruption proceeds, which may produce the Moreton waves observed in $\mathrm{H\alpha}$ wavelength~\citep{uchida1968propagation,Chen2002ApJ572L99C}. This fast-mode shock in front of the FR is invoked to produce decametric-hectometric type II coronal radio bursts associated with the CME~\citep{Magara2000ApJ538L175M,Lu2017ApJ835188L}.}

We now turn attention to the prominence formation and eruption aspects in the following sections.

  \subsection{CS evolution and prominence formation}\label{cspro}
\begin{figure*}[ht!] 
	\epsscale{1.2}
    \plotone{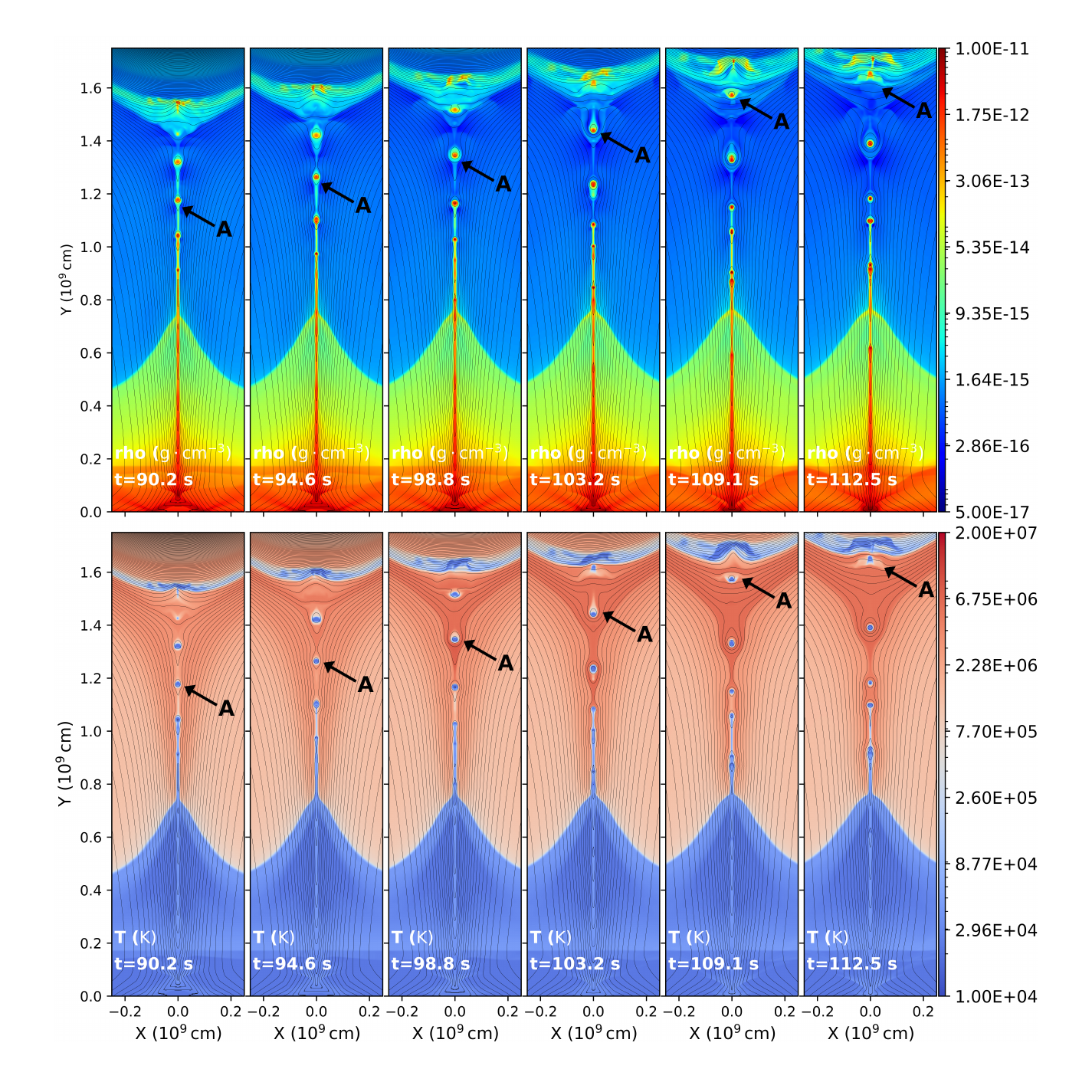}
    \caption{The evolution of mass density (top) and temperature (bottom) in the CS region with magnetic field lines overlaid (in black), illustrating the motion of magnetic islands that carry mass from the chromosphere to the FR.\label{islandE}}
\end{figure*}

The evolution of the mass density (first and second rows) and temperature (third and fourth rows) of the CS region is shown in Figure~\ref{csE} with magnetic field lines overlaid (grey).  
As the FR rises, the surrounding plasma with frozen-in field lines is driven inward underneath the FR and a CS is formed, where reconnection is then triggered. The CS formation is illustrated in details in Figure~\ref{rev}. The bottom of the FR is buried in the chromosphere at $t=0\,\mathrm{s}$ as shown in Figure~\ref{rev}. The FR is ejected upward due to catastrophe as a result of the initial non-equilibrium setup. At $t=17.2\,\mathrm{s}$, the matter from the low corona is lifted up as the FR rises while the heavy chromospheric matter tries to stay in situ and deforms the bottom of the FR.
As the top of the FR rises, the deformation of the bottom of the FR strengthens and leads to the formation of a CS at $t=20.6\,\mathrm{s}$. Some chromospheric matter is squeezed into the CS and is lifted by the concave-upward magnetic field lines that apply an upward Lorentz force on the chromospheric matter. The details and consequences of lifting chromospheric matter into the CS is here simulated: it can occur as long as the bottom of an erupting FR is partially buried in the chromosphere. We note that FR eruption driven by photospheric converging motions could also lift chromospheric plasma up to the CS height~\citep{Zhao2017,Zhao2019ApJ,Zhao2020ApJ}. A previous MHD simulation of flux emergence causing eruption~\citep{Manchester2004ApJ610} presented the similar magnetic evolution and the CS formation. In their simulation, the upper part of the FR emerged into the corona, while the lower part of the FR remained in the chromosphere/photosphere. As the upper part of the FR expanded in the corona (by shearing motion), the upper part and the lower part were gradually separated, forming the CS and new O-points. The CS underneath the FR extends in length and is thinning as the eruption proceeds as shown in the snapshot at $t=25.8\,\mathrm{s}$.
 Once the Lundquist number of the CS exceeds a critical value $S_{Lu}\simeq 10^{4}$~\citep{Bhattacharjee2009,Huang2010PhPl,Shen2011,Mei2012,Zhao2021arXiv210813508Z}, the plasmoid-mediated fast reconnection starts, accompanied by the appearance of multiple magnetic islands in the CS as shown, e.g., in the snapshot at $171.7\,\mathrm{s}$ in Figure~\ref{csE}. The time-distance diagrams of mass density (top) and temperature (bottom) along the line $x = 0$ are shown in Figure~\ref{hfr}, where the CS evolution is also illustrated. In our simulation, the plasmoid instability starts at $t\sim 40\,\mathrm{s}$, which is marked by the first vertical line in Figure~\ref{hfr}, with a Lundquist number $S_{Lu}\simeq 3.5\times 10^{4}$ of the CS. As discussed in Section~\ref{glb}, the FR first goes through an initial acceleration before $t\sim 40\,\mathrm{s}$, which coincides with the instant when the plasmoid instability starts. Before this plasmoid instability starts, some cool and dense chromospheric material has been squeezed into the CS, which is seen in earlier snapshots, e.g. at $t=34.3\,\mathrm{s}$ in Figure~\ref{csE} and snapshots in Figures~\ref{rev}. The CS regions at $t=34.3\,\mathrm{s}$ in Figure~\ref{csE} are shown in a further zoomed-in view in Figures~\ref{rev} and ~\ref{cszoom}. As also shown in Figure~\ref{hfr}, the CS is dense and cool before this time.   
The cool and dense mass in the CS is torn apart into the magnetic islands after the plasmoid instability starts. The newborn islands move upward and merge with the FR, carrying the cool and dense mass blobs into the FR. The trajectories of these islands with cool and dense matter are clearly seen in Figure~\ref{hfr} between the first and second vertical lines, during which the FR rises at a constant speed of $\sim 10^{7}\,\mathrm{cm\cdot s^{-1}}$.

The magnetic islands thus play a role of mass carriers of the chromospheric matter to the corona. This process is illustrated in Figure~\ref{islandE} in more details, where mass density (top) and temperature (bottom) in the CS region is plotted with magnetic field lines overlaid (in black). The arrows in Figure~\ref{islandE} indicate island A, which is a typical example to demonstrate how a magnetic island lifts the cool and dense remnant chromospheric matter in the CS into the FR. At $t=90.2\,\mathrm{s}$, island A, located in-between two adjacent islands, moves upward in the CS carrying dense and cool mass in it. At $t=103.2\,\mathrm{s}$, island A is reconnecting and coalescing with the FR. At $t=112.5\,\mathrm{s}$, island A is fully merged into the FR with its mass remaining at the bottom of the FR. As the plasmoid-mediated fast reconnection proceeds, more magnetic islands are produced and the coalescence of the islands with the FR repeats. The new-born magnetic islands successively push remnant chromospheric matter in the CS into the FR. These mass pieces are accumulated in the bottom of the FR, leading to the formation of dynamic blobs and threads. Hence we get a plasmoid-fed prominence formation or PF$^2$ process. After the second vertical line in Figure~\ref{hfr}, the temperature and density of the islands become closer to coronal values, and the PF$^2$ process stops. Most of the islands with cool and dense mass between the first and second vertical lines move upward, while we find almost equal fractions of downward-moving and upward-moving islands consisting of hot and tenuous matter after the second vertical line, which is consistent with the findings in~\citet{Zhao2020ApJ}. The reason why most islands with cool and dense mass move upward can be seen from Figure~\ref{rev}, e.g., at $t=17.2\,\mathrm{s}$. The bottom of the FR is deformed by the heavy cool and dense chromospheric matter and the magnetic field lines are concave-upward. The curvature of the concave-upward magnetic field lines increases as the deformation strengthens, and meanwhile the Lorentz force increases. 
The Lorentz force acting upon the cool and dense matter in the newly formed CS is directed upward. The Lorentz force gives an initial upward acceleration to the cool and dense matter, and thus most islands with cool and dense matter move upward. After the cool and dense matter move away from the CS, the magnetic field lines do not have to be concave-upward and islands consisting of hot and tenuous matter formed later move both upward and downward.

\begin{figure}[h] 
	\epsscale{1.23}
    \plotone{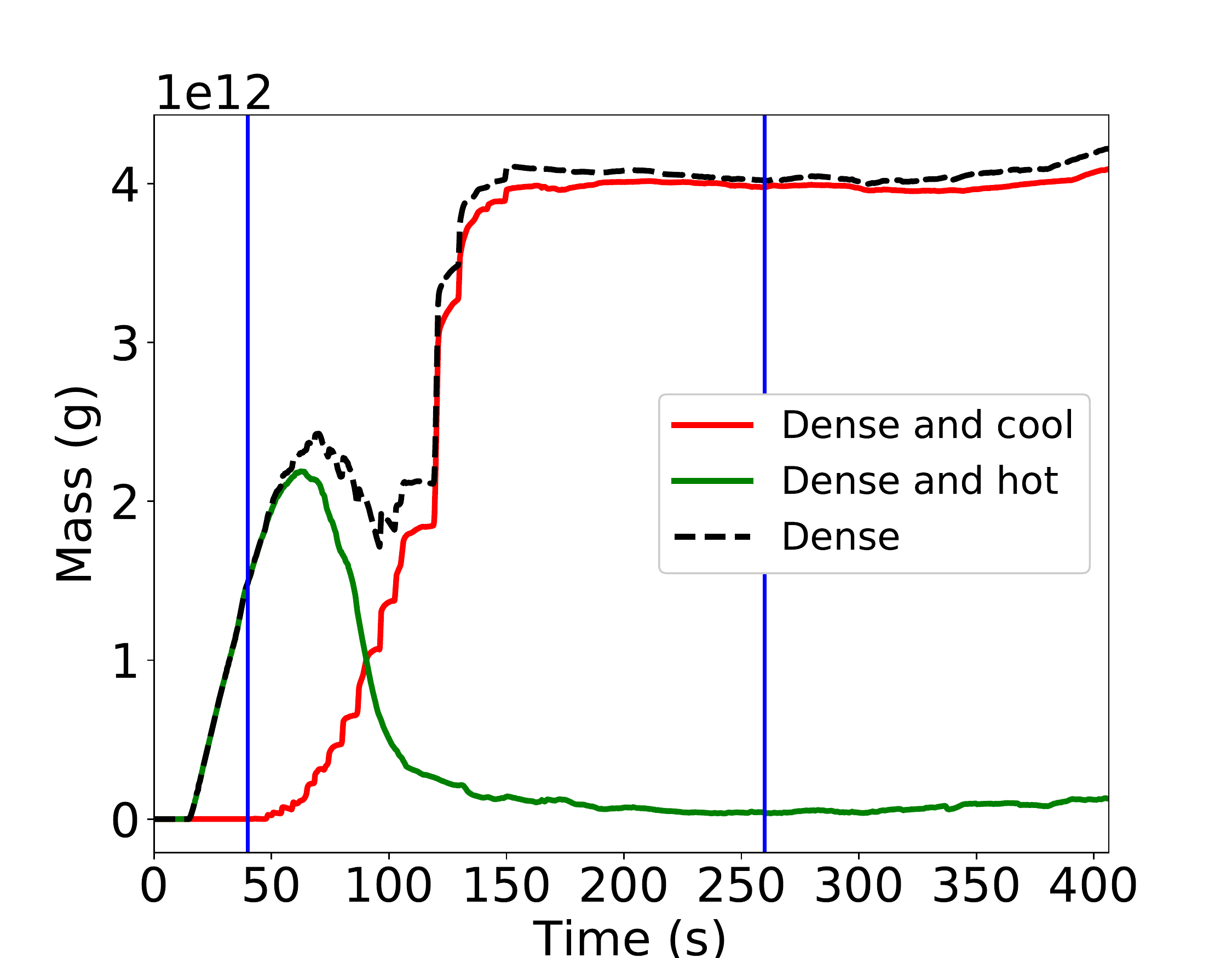}
    \caption{The time profiles of the mass of the matter that is dense and cool ($\rho>2\times 10^{-15}\,\mathrm{g\cdot cm^{-3}}$ and $T<3\times 10^{5}\,\mathrm{K}$), and dense and hot ($\rho>2\times 10^{-15}\,\mathrm{g\cdot cm^{-3}}$ and $T>3\times 10^{5}\,\mathrm{K}$) are plotted in red and green, respectively. The black dashed curve represents the mass of the dense matter at any temperature, i.e., the summation of the red and green curves.
  \label{frmass} }
\end{figure}

\begin{figure*}[ht!]
	\epsscale{1.2}
    \plotone{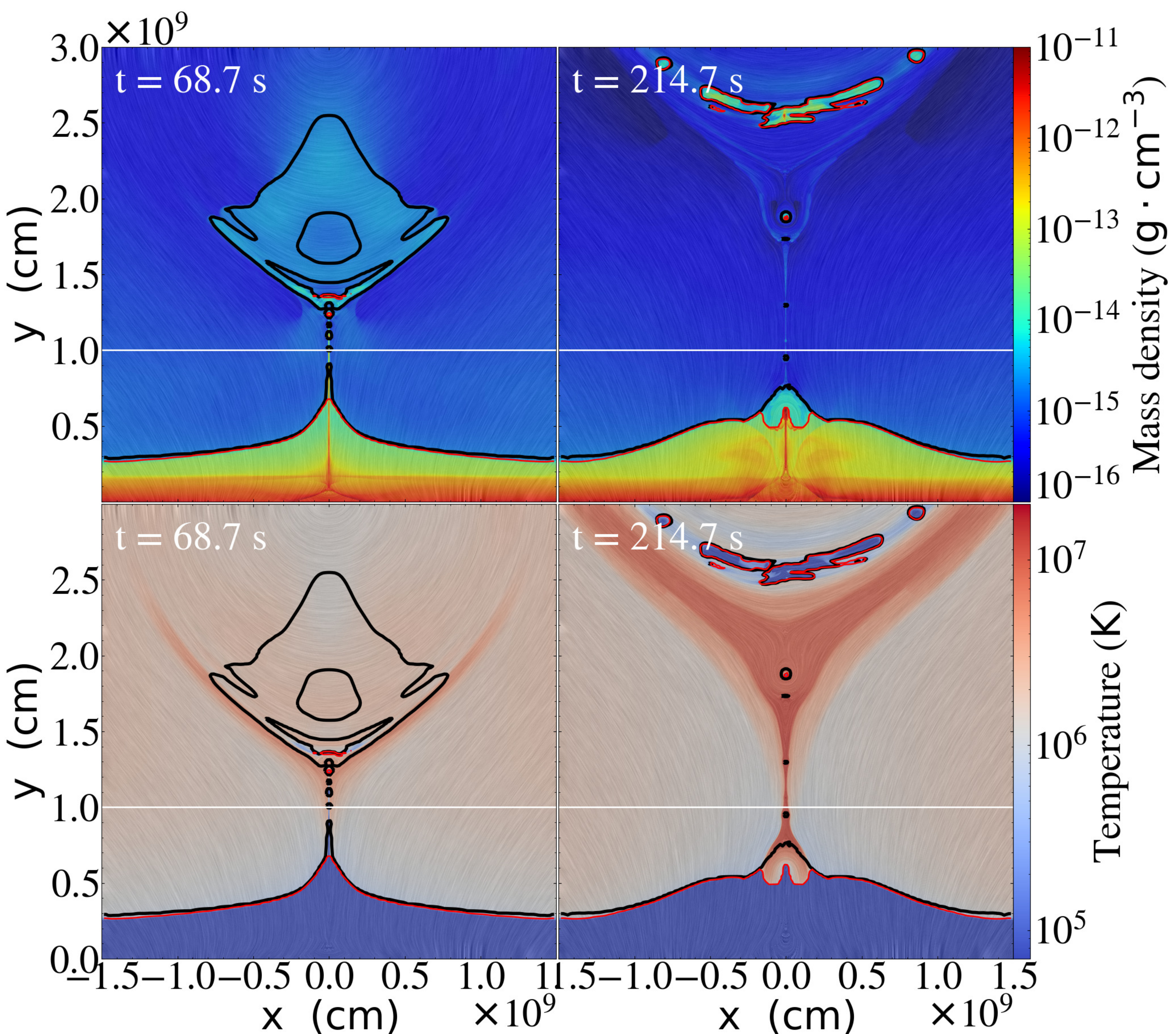}
    \caption{\label{conden} Density (top) and temperature (bottom) at $t=68.7\,\mathrm{s}$ and $t=214.7\,\mathrm{s}$, respectively, with magnetic field lines overlaid (in grey). The black contours mark 
the iso-density lines ($\rho=2\times 10^{-15}\,\mathrm{g\cdot cm^{-3}}$) and red contours represent 
the iso-temperature lines ($T=3\times 10^{5}\,\mathrm{K}$). The white lines represent the height $L_{0}=10^{9}\,\mathrm{cm}$.}
\end{figure*}

Before the plasmoid instability starts, some dense and hot material from the low corona has already been injected into the FR by the reconnection outflow and accumulated at the bottom of the FR, as shown, e.g., in the snapshots at $34.3\,\mathrm{s}$ in Figure~\ref{cszoom}. This hot and dense material cools down and condenses into the newly formed prominence as the FR rises. To illustrate the entire condensation process, we plot the time profiles of the mass of the matter that is dense and cool ($\rho>2\times 10^{-15}\,\mathrm{g\cdot cm^{-3}}$ and $T<3\times 10^{5}\,\mathrm{K}$) in red, and dense and hot ($\rho>2\times 10^{-15}\,\mathrm{g\cdot cm^{-3}}$ and $T>3\times 10^{5}\,\mathrm{K}$) in green, respectively, in Figure~\ref{frmass}. The black dashed curve represents the mass of the dense matter at any temperature, i.e., the summation of the red and green curves. Only the matter located above a height of $L_{0}=10^{9}\,\mathrm{cm}$ is counted, since dense matter below $L_{0}=10^{9}\,\mathrm{cm}$ is considered to be more closely linked to the chromosphere rather than the prominence. 
Figure~\ref{conden} shows the density (top) and temperature (bottom) at $t=68.7\,\mathrm{s}$ and $t=214.7\,\mathrm{s}$, respectively. To illustrate the dynamically evolving regions selected to calculate the mass evolution in Figure~\ref{frmass}, we plot black contours in Figure~\ref{conden} to mark 
the iso-density lines ($\rho=2\times 10^{-15}\,\mathrm{g\cdot cm^{-3}}$) and red contours to represent 
the iso-temperature lines ($T=3\times 10^{5}\,\mathrm{K}$). The white lines represent the height $L_{0}=10^{9}\,\mathrm{cm}$. The regions above the white line and encircled by the black contours are selected to calculate the mass of the dense matter in Figure~\ref{frmass}. The intersections of the regions encircled by the red and black contours and located above the white line constitute the dense and cool matter.
Due to the two-and-a-half-dimensional setup, our system is translationally invariant in the {\it z}-direction. Therefore, to calculate the total prominence mass by an $(x,y)$-masked, 3D integral $\int_{\mathrm{mask}} \rho(x,y,t)\mathrm{d}x\mathrm{d}y\mathrm{d}z$, we artificially take the ignored {\it z}-direction as $[0, L_{0}]$, i.e., the prominence is supposed to have a length of $L_{0}=10^{9}\,\mathrm{cm}$.  
 The red cool-and-dense curve starts to increase from $t\sim 50 \,\mathrm{s}$, $10 \,\mathrm{s}$ after the plasmoid instability starts, which indicates that some dense and cool chromospheric matter is lifted above $L_{0}=10^{9}\,\mathrm{cm}$ by the magnetic islands. The green curve starts to increase from an earlier instant of time $t\sim 20 \,\mathrm{s}$ because the dense and hot matter exists in a higher region than the chromospheric matter and is thus lifted above the height $L_{0}=10^{9}\,\mathrm{cm}$ by the FR at an earlier time. The green curve decreases from $t\sim 68.7\,\mathrm{s}$, a few seconds after the red curve increases, which marks the beginning of the condensation.  
The dense and cool matter accumulated at the bottom of the FR is about $4\times 10^{11}\,\mathrm{g}$ when the condensation begins. The dense and cool matter perturbs the density and temperature distributions, triggering thermal instabilities. The hot and dense matter at the bottom of the FR continuously condenses into the newly formed prominence. The condensation process is more vividly depicted in Figure~\ref{conden}. The dense matter encircled by the black contours in an earlier instant $t=68.7\,\mathrm{s}$ is clearly much larger than the region encircled by the red contours (i.e., the prominence). The black contours gradually contract and finally overlay with the red contours that encircle the cool matter in the later instant $t=214.7\,\mathrm{s}$. Meanwhile, the area encircled by the red contours also grows from $t=68.7\,\mathrm{s}$ to $t=214.7\,\mathrm{s}$. This corresponds in Figure~\ref{frmass} to the black curve being much higher than the red curve at $t=68.7\,\mathrm{s}$, but the two curves almost overlay with each other at $t=214.7\,\mathrm{s}$, and the red curve rises from $t=68.7\,\mathrm{s}$ to $t=214.7\,\mathrm{s}$. 
 The condensation mainly occurs in the bottom of the FR and lasts from $t\sim  68.7\,\mathrm{s}$ to $t\sim  150 \,\mathrm{s}$, which falls into the time interval between the two vertical lines in Figure~\ref{frmass} that mark the beginning and end of the second phase of the FR eruption. Afterwards, the mass of the dense and hot matter (green curve) is ignorable and the red and black curves almost overlay with each other. 
 Before the condensation starts from $t\sim  68.7\,\mathrm{s}$, the prominence (the dense and cool matter) formation is driven purely by the newly discovered PF$^2$ process. The mass of the prominence at $t\sim  68.7\,\mathrm{s}$ is $\sim 3\times10^{11}\,\mathrm{g}$. After the condensation, the mass of the prominence reaches $\sim 4\times 10^{12}\,\mathrm{g}$ at $t\sim  150\,\mathrm{s}$.
 Although, the PF$^2$ process lasts until $t\sim 260 \,\mathrm{s}$, the prominence mass keeps almost constant $\sim 4\times 10^{12}\,\mathrm{g}$ after $t\sim 150 \,\mathrm{s}$ because only very few islands carry cool and dense matter to the FR and the mass gained from the PF$^2$ process after $t\sim 150 \,\mathrm{s}$ is only $\sim 1\times10^{11}\,\mathrm{g}$. The total mass gained from the condensation process is $\sim 2\times 10^{12}\,\mathrm{g}$, which is seen from the peak of the green curve, while the other half of the mass $\sim 2\times 10^{12}\,\mathrm{g}$ is from the PF$^2$ process. In early literature, \citet{kuperus1967nature} also suggested that condensation might take place in the CS and plasmoids formed by resistive instabilities would cause a filamentary thread structure in the forming prominence, as generally observed. However, their (cartoon) model is an in-situ condensation model for quiescent prominences, where the prominence mass comes from the corona by condensation. The mass transfer from the chromosphere to the corona by plasmoids, i.e., the PF$^2$ process is a new ingredient to explain eruptive prominence formations in our work.

\subsection{Forward modeling}\label{Forward} 

\begin{figure*}[ht!]
	\epsscale{1.1}
    \plotone{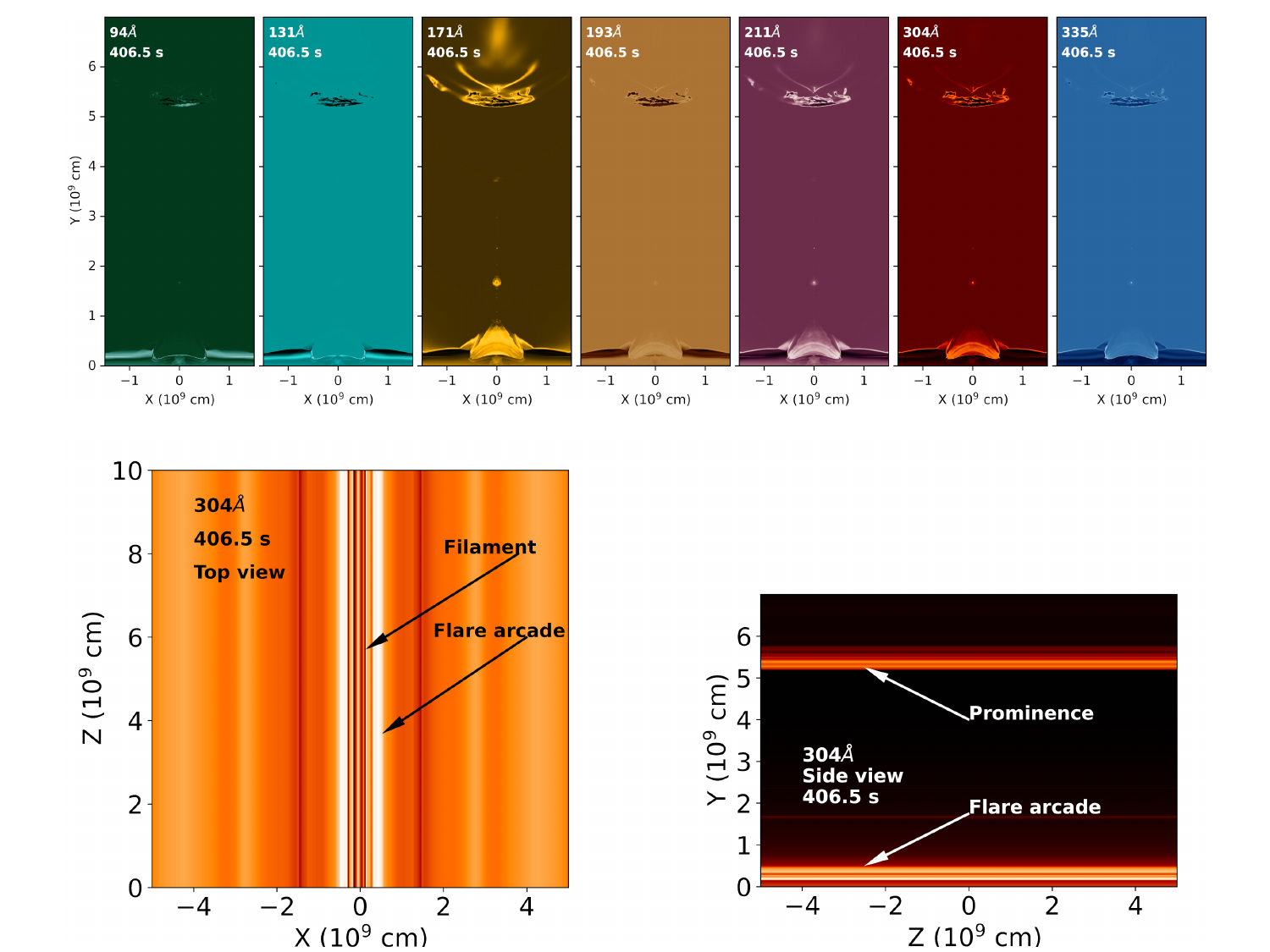}
    \caption{\label{forward} The upper panels show the synthesized SDO/AIA $94\AA$, $131\AA$, $171\AA$, $193\AA$, $211\AA$, $304\AA$, and $335\AA$ images in the {\it x-y} plane at $406.5\,\mathrm{s}$. The bottom-left panel shows the synthesized SDO/AIA $304\AA$ image in the {\it x-z} plane viewed from the top and the image in the {\it z-y} plane viewed from the side is shown in the bottom-right panel.}
\end{figure*} 

To compare our numerical results with observations, synthesized extreme ultraviolet (EUV) radiations of the seven channels of SDO/AIA, i.e., $94\AA$, $131\AA$, $171\AA$, $193\AA$, $211\AA$, $304\AA$, and $335\AA$, are calculated based on the temperature and density obtained from the MHD simulation by forward modeling analysis. The main task of the forward modeling analysis is to calculate the radiation intensity $I_{\lambda}$ by solving the radiative transfer equation~\citep{Rybicki1986,Zhao2019ApJ}:
\begin{equation}
I_{\lambda}(\tau_{\lambda})=I_{\lambda}(0)\exp(-\tau_{\lambda}) +\int_{0}^{\tau_{\lambda}}\exp[-(\tau_{\lambda}-\tau_{\lambda}^{\prime})] S_{\lambda}(\tau_{\lambda}^{\prime})\mathrm{d}\tau_{\lambda}^{\prime},\label{eq6}
\end{equation}
where $S_{\lambda}$ is the source function and $\tau_{\lambda}$ is the optical depth. 
To calculate the radiation in the {\it x-y} plane, the line-of-sight integral in the radiative transfer equation is taken along the {\it z}-axis in the interval $[-5L_{0},5L_{0}]$, along which the source function $S_{\lambda}$ is constant because the system is translationally invariant due to the two-and-a-half-dimensional setup, while, in the {\it x-z} and {\it z-y} planes, i.e., the top and side views, the line-of-sight integrals are taken in the {\it y}- and {\it x}-directions in the intervals $[0,7L_{0}]$ and $[-5L_{0},5L_{0}]$, respectively, where the source function $S_{\lambda}$ is not a constant anymore. Here $L_{0}=10^{9}\,\mathrm{cm}$ is the typical length scale as listed in Table~\ref{Units}. The background radiation $I_{\lambda}(0)$ is taken as $1\%$ of the highest radiation value. The main technical details of our forward modeling analysis are fully described in~\citet{Zhao2019ApJ}.

Synthetic EUV images at $406.5\,\mathrm{s}$ of the seven channels of SDO/AIA, in the $(x,y)$ view are plotted in the top panels in Figure~\ref{forward}. The plasmoids are visible in the $171\AA$, $211\AA$, $304\AA$, and $335\AA$ channels, and are especially apparent in the $171\AA$ channel, which are easily identified as bright blobs. The flare loops appear bright in the $171\AA$, $211\AA$, and $304\AA$ channels. We can also identify fibril-like thread structure of the prominence in all the seven channels.

The synthetic EUV $304\AA$ images in the {\it x-z} and {\it z-y} planes, i.e., the top and side views, at the same time are also calculated and shown in the bottom-left and bottom-right panels of Figure~\ref{forward}, respectively. The bright flare arcade and the filament threads are clearly visible in the top and side views, as indicated by the arrows. Due to the translational invariance along the {\it z}-direction in our two-and-a-half-dimensional setup, all the filament threads are exactly oriented along the {\it z}-direction, and manifest themselves as elongated fibrils above the underlying polarity inversion line. In the recent two-dimensional simulations of an arcade-supported prominence in~\citet{zhou2020simulations}, some filament threads are found to be misaligned with the local magnetic field by $\sim 2^{\circ}$.

\section{Summary and conclusion}\label{conclusion}

In this study, we demonstrate prominence formation during the FR eruption process, in a chromosphere-transition region-corona setup with gravity, resistivity, thermal conduction, coronal background heating, and optically-thin radiative cooling effects included. The initial setup gives a global Lundquist number $S_{Lu}\simeq 5\times 10^{5}$. By using grid-adaptivity, our simulation covers a wide range of scales, from the large-scale CME kinematics to the meso-scale magnetic island dynamics. In contrast with our previous levitation model, an FR pre-exists in the lower corona in this model. This pre-existing FR erupts due to catastrophe, with a CS formed underneath. The erupting FR drives the formation of a fast-mode shock in front of it. Some chromospheric matter is squeezed into the CS during the sheet formation. The plasmoid instability occurs in the CS when the Lundquist number of the CS reaches $\sim 3.5\times10^{4}$. These plasmoids take remnant chromospheric matter from the CS into the FR, leading to the formation of a prominence by a new plasmoid-fed prominence formation or PF$^2$ process. Moreover, hot and dense plasma levitated from the low corona by the FR condenses into the prominence as the FR rises. Our model naturally produces fragmented mass threads, rather than a vertical “sheet-like” prominence arising in the evaporation-condensation model of~\citet{Xia2011}.
Synthetic EUV images of the seven channels of SDO/AIA, i.e., $94\AA$, $131\AA$, $171\AA$, $193\AA$, $211\AA$, $304\AA$, and $335\AA$ are reproduced by forward modeling analysis, and the fragmented mass threads are clearly seen in EUV images. Our simulation also captures many details relevant to a complete understanding of FR/prominence eruptions.
This study can thus act as a starting point for future studies of 2.5 or 3D CME dynamics. We can e.g. study particle acceleration aspects in the fast-mode shock in front of the CME and in the CS underneath.

\appendix
A two-and-a-half-dimensional magnetic field $\mathbf{B}(x,y)$ can be represented as $\mathbf{B}=\nabla\times (A\mathbf{e}_{z})+B_{z}\mathbf{e}_{z}$, where $A(x,y)$ is the magnetic flux function. We adopt a magnetic field such that the magnetic flux function $A(x,y)$ can be expressed as the sum of three contributions as follows:
\begin{equation}
	A(x,y)=A_{\mathrm{FR}}(x,y)+A_{\mathrm{Im}}(x,y)+A_{q}(x,y),
\end{equation}
where $A_{\mathrm{FR}}$ is the magnetic flux by the FR current in the corona, $A_{\mathrm{Im}}$ is the magnetic flux by the image current of the
FR beneath the photosphere, and $A_{q}$ is the magnetic flux of a magnetic quadrupole.

The magnetic flux $A_{\mathrm{FR}}$ by the FR current in the corona is defined as follows
\begin{equation}
 A_{\mathrm{FR}} =\left \{ 
 \begin{split}
 	& A_{0}R_{0}\left[\left(\frac{r_{1}}{R_{0}}\right)^{2}-\frac{1}{4}\left(\frac{r_{1}}{R_{0}}\right)^{4}-\frac{3}{2}\right],r_{1}<R_{0}\\&
 	-A_{0}R_{0}\left[\frac{3}{4}-\ln\left(\frac{r_{1}}{R_{0}}\right) \right],r_{1}\ge R_{0}
 	 \end{split}
  \right.,
\end{equation}
where $R_0$ is the radius of the flux rope and $A_0$ specifies the magnetic field strength. Here $r_{1}$ is the distance from any point $(x,y)$ in the upper {\it x-y} plane to the FR centre at $(0,H+2\delta)$, so $r_1(x,y)$ is defined as 
\begin{equation}
	r_{1}=\sqrt{x^{2}+[y-(H+2\delta)]^{2}},
\end{equation}
where $(H+2\delta)$ is the height of the centre of the FR, and $\delta$ is a small parameter to adjust the height of the FR.

The magnetic flux $A_{\mathrm{Im}}$ by the image current of the FR beneath the photosphere is 
\begin{equation}
	A_{\mathrm{Im}}=-A_{0}R_{0}\ln\left(\frac{r_{2}}{R_{0}}\right),
\end{equation}
where the distance $r_2$ from any point $(x,y)$ to the centre of the image FR at $(0, -H + 2\delta)$ beneath the photosphere is
\begin{equation}
	r_{2}=\sqrt{x^{2}+[y-(-H+2\delta)]^{2}}.
\end{equation}

The magnetic flux of the magnetic quadrupole
buried at the depth $y_{d}=-D+2\delta$ is
\begin{equation}
	A_{q}=\frac{A_{0}R_{0}D^{2}M_{q}}{2 r^{4}_{3}}[x^{2}-(y+D-2\delta)^{2}].
\end{equation}
Here $r_{3}=\sqrt{x^{2}+[y-(-D+2\delta)]^{2}}$, $D=2H$, and $M_{q}$ is a parameter controlling the force balance of the FR. When $M_{q}=27/8$, the FR is in an equilibrium state. When $M_{q}$ is smaller than the critical value $M_{c}=27/8$, no equilibrium can be found in this case and the FR will be ejected upward due to loss of equilibrium~\citep{Takahashi2017}.

Finally, the in-plane {\it z}-component of the magnetic field $B_{z}(x,y)$ is given by
\begin{equation}
 B_{z} =\left \{ 
 \begin{split}
 	& A_{0}\sqrt{ \frac{2}{3}\left[5-12\left(\frac{r_{1}}{R_{0}}\right)^{2}+9\left(\frac{r_{1}}{R_{0}}\right)^{4} -2\left(\frac{r_{1}}{R_{0}}\right)^{6}\right]},\,\,\, r_{1}<R_{0} \,,\\&
 	0, \,\,\, r_{1}\ge R_{0}\,.
 	 \end{split}
  \right..
\end{equation}
 
\begin{acknowledgments}
We acknowledge support by a joint FWO-NSFC grant G0E9619N and by FWO project G0B4521N. RK received funding from the European Research Council (ERC) under the European Unions Horizon 2020 research and innovation programme (grant agreement No. 833251 PROMINENT ERC-ADG 2018), and from Internal Funds KU Leuven, project C14/19/089 TRACESpace.
\end{acknowledgments}

\bibliographystyle{apj}
\bibliography{reference}

\begin{thebibliography}{}
\expandafter\ifx\csname natexlab\endcsname\relax\def\natexlab#1{#1}\fi
\providecommand{\url}[1]{\href{#1}{#1}}
\providecommand{\dodoi}[1]{doi:~\href{http://doi.org/#1}{\nolinkurl{#1}}}
\providecommand{\doeprint}[1]{\href{http://ascl.net/#1}{\nolinkurl{http://ascl.net/#1}}}
\providecommand{\doarXiv}[1]{\href{https://arxiv.org/abs/#1}{\nolinkurl{https://arxiv.org/abs/#1}}}

\bibitem[{{Aly}(1990)}]{Aly1990}
{Aly}, J.~J. 1990, Computer Physics Communications, 59, 13,
  \dodoi{10.1016/0010-4655(90)90152-Q}

\bibitem[{An {et~al.}(1988)An, Bao, Wu, \& Suess}]{an1988numerical}
An, C.-H., Bao, J., Wu, S., \& Suess, S. 1988, Solar physics, 115, 93

\bibitem[{{Antiochos} {et~al.}(1999){Antiochos}, {DeVore}, \&
  {Klimchuk}}]{Antiochos1999}
{Antiochos}, S.~K., {DeVore}, C.~R., \& {Klimchuk}, J.~A. 1999, \apj, 510, 485,
  \dodoi{10.1086/306563}

\bibitem[{Antiochos {et~al.}(1999)Antiochos, MacNeice, Spicer, \&
  Klimchuk}]{antiochos1999dynamic}
Antiochos, S.~K., MacNeice, P.~J., Spicer, D.~S., \& Klimchuk, J.~A. 1999, The
  Astrophysical Journal, 512, 985

\bibitem[{{B{\'a}rta} {et~al.}(2011){B{\'a}rta}, {B{\"u}chner}, {Karlick{\'y}},
  \& {Sk{\'a}la}}]{Barta2011}
{B{\'a}rta}, M., {B{\"u}chner}, J., {Karlick{\'y}}, M., \& {Sk{\'a}la}, J.
  2011, \apj, 737, 24, \dodoi{10.1088/0004-637X/737/1/24}

\bibitem[{{Bhattacharjee} {et~al.}(2009){Bhattacharjee}, {Huang}, {Yang}, \&
  {Rogers}}]{Bhattacharjee2009}
{Bhattacharjee}, A., {Huang}, Y.-M., {Yang}, H., \& {Rogers}, B. 2009, Physics
  of Plasmas, 16, 112102, \dodoi{10.1063/1.3264103}

\bibitem[{{Carmichael}(1964)}]{Carmichael1964}
{Carmichael}, H. 1964, NASA Special Publication, 50, 451

\bibitem[{{Chen}(2011)}]{Chen2011}
{Chen}, P.~F. 2011, Living Reviews in Solar Physics, 8, 1,
  \dodoi{10.12942/lrsp-2011-1}

\bibitem[{{Chen} \& {Shibata}(2000)}]{Chen2000}
{Chen}, P.~F., \& {Shibata}, K. 2000, \apj, 545, 524, \dodoi{10.1086/317803}

\bibitem[{Dungey(1953)}]{dungey1953lxxvi}
Dungey, J. 1953, The London, Edinburgh, and Dublin Philosophical Magazine and
  Journal of Science, 44, 725

\bibitem[{Fan(2018)}]{fan2018mhd}
Fan, Y. 2018, The Astrophysical Journal, 862, 54

\bibitem[{{Forbes} \& {Isenberg}(1991)}]{ForbesIsenberg1991}
{Forbes}, T.~G., \& {Isenberg}, P.~A. 1991, \apj, 373, 294,
  \dodoi{10.1086/170051}

\bibitem[{Guo {et~al.}(2019)Guo, DF, LH, {et~al.}}]{guo2019formation}
Guo, Q., DF, K., LH, Y., {et~al.} 2019

\bibitem[{{Hirayama}(1974)}]{Hirayama1974}
{Hirayama}, T. 1974, \solphys, 34, 323, \dodoi{10.1007/BF00153671}

\bibitem[{{Huang} \& {Bhattacharjee}(2010)}]{Huang2010PhPl}
{Huang}, Y.-M., \& {Bhattacharjee}, A. 2010, Physics of Plasmas, 17, 062104,
  \dodoi{10.1063/1.3420208}

\bibitem[{{Illing} \& {Hundhausen}(1983)}]{Illing1983}
{Illing}, R.~M.~E., \& {Hundhausen}, A.~J. 1983, \jgr, 88, 10210,
  \dodoi{10.1029/JA088iA12p10210}

\bibitem[{{Illing} \& {Hundhausen}(1985)}]{Illing1985}
---. 1985, \jgr, 90, 275, \dodoi{10.1029/JA090iA01p00275}

\bibitem[{Jenkins \& Keppens(2021)}]{jenkins2021prominence}
Jenkins, J., \& Keppens, R. 2021, Astronomy \& Astrophysics, 646, A134

\bibitem[{Kaneko \& Yokoyama(2017)}]{kaneko2017reconnection}
Kaneko, T., \& Yokoyama, T. 2017, The Astrophysical Journal, 845, 12

\bibitem[{Keppens {et~al.}(2019)Keppens, Guo, Makwana, Mei, Ripperda, Xia, \&
  Zhao}]{keppens2019ideal}
Keppens, R., Guo, Y., Makwana, K., {et~al.} 2019, Reviews of Modern Plasma
  Physics, 3, 14

\bibitem[{{Keppens} {et~al.}(2014){Keppens}, {Porth}, \&
  {Xia}}]{Keppens2014ApJ77K}
{Keppens}, R., {Porth}, O., \& {Xia}, C. 2014, \apj, 795, 77,
  \dodoi{10.1088/0004-637X/795/1/77}

\bibitem[{Keppens {et~al.}(2021)Keppens, Teunissen, Xia, \&
  Porth}]{keppens2021mpi}
Keppens, R., Teunissen, J., Xia, C., \& Porth, O. 2021, Computers \&
  Mathematics with Applications, 81, 316

\bibitem[{{Ko} {et~al.}(2003){Ko}, {Raymond}, {Lin}, {Lawrence}, {Li}, \&
  {Fludra}}]{Ko2003ApJ}
{Ko}, Y.-K., {Raymond}, J.~C., {Lin}, J., {et~al.} 2003, \apj, 594, 1068,
  \dodoi{10.1086/376982}

\bibitem[{{Kopp} \& {Pneuman}(1976)}]{KoppPneuman1976}
{Kopp}, R.~A., \& {Pneuman}, G.~W. 1976, \solphys, 50, 85,
  \dodoi{10.1007/BF00206193}

\bibitem[{Kuperus \& Tandberg-Hanssen(1967)}]{kuperus1967nature}
Kuperus, M., \& Tandberg-Hanssen, E. 1967, Solar Physics, 2, 39

\bibitem[{Leroy(1989)}]{leroy1989observation}
Leroy, J. 1989, Observation of Prominence Magnetic Fields, Dynamics and
  Structures of Quiescent Prominences, ed. ER Priest,  Kluwer Academic
  Publishers, Dordrecht, Holland

\bibitem[{{Lin} \& {Forbes}(2000)}]{LinForbes2000}
{Lin}, J., \& {Forbes}, T.~G. 2000, \jgr, 105, 2375,
  \dodoi{10.1029/1999JA900477}

\bibitem[{{Lu} {et~al.}(2017){Lu}, {Inhester}, {Feng}, {Liu}, \&
  {Zhao}}]{Lu2017ApJ835188L}
{Lu}, L., {Inhester}, B., {Feng}, L., {Liu}, S., \& {Zhao}, X. 2017, \apj, 835,
  188, \dodoi{10.3847/1538-4357/835/2/188}

\bibitem[{Mackay {et~al.}(2010)Mackay, Karpen, Ballester, Schmieder, \&
  Aulanier}]{mackay2010physics}
Mackay, D., Karpen, J., Ballester, J., Schmieder, B., \& Aulanier, G. 2010,
  Space Science Reviews, 151, 333

\bibitem[{{Manchester} {et~al.}(2004){Manchester}, {Gombosi}, {DeZeeuw}, \&
  {Fan}}]{Manchester2004ApJ610}
{Manchester}, W., I., {Gombosi}, T., {DeZeeuw}, D., \& {Fan}, Y. 2004, \apj,
  610, 588, \dodoi{10.1086/421516}

\bibitem[{Martens \& Kuin(1989)}]{martens1989circuit}
Martens, P., \& Kuin, N. 1989, Solar physics, 122, 263

\bibitem[{{Mei} {et~al.}(2012){Mei}, {Shen}, {Wu}, {Lin}, {Murphy}, \&
  {Roussev}}]{Mei2012}
{Mei}, Z., {Shen}, C., {Wu}, N., {et~al.} 2012, \mnras, 425, 2824,
  \dodoi{10.1111/j.1365-2966.2012.21625.x}

\bibitem[{{Ni} \& {Lukin}(2018)}]{Ni2018ApJ}
{Ni}, L., \& {Lukin}, V.~S. 2018, \apj, 868, 144,
  \dodoi{10.3847/1538-4357/aaeb97}

\bibitem[{{Ni} {et~al.}(2012){Ni}, {Roussev}, {Lin}, \&
  {Ziegler}}]{Ni2012ApJ75820N}
{Ni}, L., {Roussev}, I.~I., {Lin}, J., \& {Ziegler}, U. 2012, \apj, 758, 20,
  \dodoi{10.1088/0004-637X/758/1/20}

\bibitem[{Nool \& Keppens(2002)}]{Nool2002}
Nool, M., \& Keppens, R. 2002, Computational Methods in Applied Mathematics, 2,
  92

\bibitem[{{Parker}(1957)}]{Parker1957}
{Parker}, E.~N. 1957, \jgr, 62, 509, \dodoi{10.1029/JZ062i004p00509}

\bibitem[{Parker(1963)}]{parker1963solar}
Parker, E.~N. 1963, The Astrophysical Journal Supplement Series, 8, 177

\bibitem[{{Petschek}(1964)}]{Petschek1964}
{Petschek}, H.~E. 1964, NASA Special Publication, 50, 425

\bibitem[{{Porth} {et~al.}(2014){Porth}, {Xia}, {Hendrix}, {Moschou}, \&
  {Keppens}}]{Porth2014ApJS}
{Porth}, O., {Xia}, C., {Hendrix}, T., {Moschou}, S.~P., \& {Keppens}, R. 2014,
  \apjs, 214, 4, \dodoi{10.1088/0067-0049/214/1/4}

\bibitem[{{Qiu} {et~al.}(2002){Qiu}, {Lee}, {Gary}, \&
  {Wang}}]{Qiu2002ApJ5651335Q}
{Qiu}, J., {Lee}, J., {Gary}, D.~E., \& {Wang}, H. 2002, \apj, 565, 1335,
  \dodoi{10.1086/324706}

\bibitem[{{Reeves} {et~al.}(2010){Reeves}, {Linker}, {Miki{\'c}}, \&
  {Forbes}}]{Reeves2010ApJ72}
{Reeves}, K.~K., {Linker}, J.~A., {Miki{\'c}}, Z., \& {Forbes}, T.~G. 2010,
  \apj, 721, 1547, \dodoi{10.1088/0004-637X/721/2/1547}

\bibitem[{{Ruan} {et~al.}(2020){Ruan}, {Xia}, \& {Keppens}}]{Ruan2020ApJ}
{Ruan}, W., {Xia}, C., \& {Keppens}, R. 2020, \apj, 896, 97,
  \dodoi{10.3847/1538-4357/ab93db}

\bibitem[{{Rybicki} \& {Lightman}(1986)}]{Rybicki1986}
{Rybicki}, G.~B., \& {Lightman}, A.~P. 1986, {Radiative Processes in
  Astrophysics}, 400

\bibitem[{{Shen} {et~al.}(2011){Shen}, {Lin}, \& {Murphy}}]{Shen2011}
{Shen}, C., {Lin}, J., \& {Murphy}, N.~A. 2011, \apj, 737, 14,
  \dodoi{10.1088/0004-637X/737/1/14}

\bibitem[{{Shibata} \& {Tanuma}(2001)}]{ShibataTanuma2001}
{Shibata}, K., \& {Tanuma}, S. 2001, Earth, Planets, and Space, 53, 473,
  \dodoi{10.1186/BF03353258}

\bibitem[{Spiteri \& Ruuth(2002)}]{spiteri2002new}
Spiteri, R.~J., \& Ruuth, S.~J. 2002, SIAM Journal on Numerical Analysis, 40,
  469

\bibitem[{{Sturrock}(1966)}]{Sturrock1966}
{Sturrock}, P.~A. 1966, \nat, 211, 695, \dodoi{10.1038/211695a0}

\bibitem[{{Sweet}(1958{\natexlab{a}})}]{sweet1958production}
{Sweet}, P.~A. 1958{\natexlab{a}}, Il Nuovo Cimento (1955-1965), 8, 188

\bibitem[{{Sweet}(1958{\natexlab{b}})}]{Sweet1958}
{Sweet}, P.~A. 1958{\natexlab{b}}, in IAU Symposium, Vol.~6, Electromagnetic
  Phenomena in Cosmical Physics, ed. B.~{Lehnert}, 123

\bibitem[{{Takahashi} {et~al.}(2017){Takahashi}, {Qiu}, \&
  {Shibata}}]{Takahashi2017}
{Takahashi}, T., {Qiu}, J., \& {Shibata}, K. 2017, \apj, 848, 102,
  \dodoi{10.3847/1538-4357/aa8f97}

\bibitem[{{Titov} \& {D{\'e}moulin}(1999)}]{Titov1999AA}
{Titov}, V.~S., \& {D{\'e}moulin}, P. 1999, \aap, 351, 707

\bibitem[{{T{\"o}r{\"o}k} \& {Kliem}(2005)}]{Torok2005ApJ}
{T{\"o}r{\"o}k}, T., \& {Kliem}, B. 2005, \apjl, 630, L97,
  \dodoi{10.1086/462412}

\bibitem[{{T{\'o}th} \& {Odstr{\v{c}}il}(1996)}]{Toth1996JCoPh}
{T{\'o}th}, G., \& {Odstr{\v{c}}il}, D. 1996, Journal of Computational Physics,
  128, 82, \dodoi{10.1006/jcph.1996.0197}

\bibitem[{{{\v{C}}ada} \& {Torrilhon}(2009)}]{Cada2009JCoPh}
{{\v{C}}ada}, M., \& {Torrilhon}, M. 2009, Journal of Computational Physics,
  228, 4118, \dodoi{10.1016/j.jcp.2009.02.020}

\bibitem[{Wang(1999)}]{wang1999jetlike}
Wang, Y.-M. 1999, The Astrophysical Journal Letters, 520, L71

\bibitem[{{Webb} \& {Howard}(2012)}]{Webb2012}
{Webb}, D.~F., \& {Howard}, T.~A. 2012, Living Reviews in Solar Physics, 9, 3,
  \dodoi{10.12942/lrsp-2012-3}

\bibitem[{{Xia} {et~al.}(2012){Xia}, {Chen}, \& {Keppens}}]{Xia2012}
{Xia}, C., {Chen}, P.~F., \& {Keppens}, R. 2012, \apjl, 748, L26,
  \dodoi{10.1088/2041-8205/748/2/L26}

\bibitem[{{Xia} {et~al.}(2011){Xia}, {Chen}, {Keppens}, \& {van
  Marle}}]{Xia2011}
{Xia}, C., {Chen}, P.~F., {Keppens}, R., \& {van Marle}, A.~J. 2011, \apj, 737,
  27, \dodoi{10.1088/0004-637X/737/1/27}

\bibitem[{{Xia} \& {Keppens}(2016)}]{Xia2016}
{Xia}, C., \& {Keppens}, R. 2016, \apj, 823, 22,
  \dodoi{10.3847/0004-637X/823/1/22}

\bibitem[{{Xia} {et~al.}(2014){Xia}, {Keppens}, {Antolin}, \&
  {Porth}}]{Xia2014ApJ}
{Xia}, C., {Keppens}, R., {Antolin}, P., \& {Porth}, O. 2014, \apjl, 792, L38,
  \dodoi{10.1088/2041-8205/792/2/L38}

\bibitem[{{Xia} {et~al.}(2018){Xia}, {Teunissen}, {El Mellah}, {Chan{\'e}}, \&
  {Keppens}}]{Xia2018ApJS}
{Xia}, C., {Teunissen}, J., {El Mellah}, I., {Chan{\'e}}, E., \& {Keppens}, R.
  2018, \apjs, 234, 30, \dodoi{10.3847/1538-4365/aaa6c8}

\bibitem[{{Ye} {et~al.}(2020){Ye}, {Cai}, {Shen}, {Raymond}, {Lin}, {Roussev},
  \& {Mei}}]{Ye2020ApJ89764Y}
{Ye}, J., {Cai}, Q., {Shen}, C., {et~al.} 2020, \apj, 897, 64,
  \dodoi{10.3847/1538-4357/ab93b5}

\bibitem[{{Ye} {et~al.}(2021){Ye}, {Cai}, {Shen}, {Raymond}, {Mei}, {Li}, \&
  {Lin}}]{Ye2021ApJ90945Y}
---. 2021, \apj, 909, 45, \dodoi{10.3847/1538-4357/abdeb5}

\bibitem[{{Ye} {et~al.}(2017){Ye}, {Lin}, {Raymond}, \&
  {Shen}}]{Ye2017AGUFMSH11B2436Y}
{Ye}, J., {Lin}, J., {Raymond}, J.~C., \& {Shen}, C. 2017, in AGU Fall Meeting
  Abstracts, Vol. 2017, SH11B--2436

\bibitem[{Yee(1989)}]{yee1989class}
Yee, H.~C. 1989

\bibitem[{{Zhao} {et~al.}(2021){Zhao}, {Bacchini}, \&
  {Keppens}}]{Zhao2021arXiv210813508Z}
{Zhao}, X., {Bacchini}, F., \& {Keppens}, R. 2021, Physics of Plasmas, 28,
  092113, \dodoi{10.1063/5.0058326}

\bibitem[{{Zhao} \& {Keppens}(2020)}]{Zhao2020ApJ}
{Zhao}, X., \& {Keppens}, R. 2020, \apj, 898, 90,
  \dodoi{10.3847/1538-4357/ab9a31}

\bibitem[{{Zhao} {et~al.}(2017){Zhao}, {Xia}, {Keppens}, \& {Gan}}]{Zhao2017}
{Zhao}, X., {Xia}, C., {Keppens}, R., \& {Gan}, W. 2017, \apj, 841, 106,
  \dodoi{10.3847/1538-4357/aa7142}

\bibitem[{{Zhao} {et~al.}(2019){Zhao}, {Xia}, {Van Doorsselaere}, {Keppens}, \&
  {Gan}}]{Zhao2019ApJ}
{Zhao}, X., {Xia}, C., {Van Doorsselaere}, T., {Keppens}, R., \& {Gan}, W.
  2019, \apj, 872, 190, \dodoi{10.3847/1538-4357/ab0284}

\bibitem[{Zhou {et~al.}(2020)Zhou, Chen, Hong, \& Fang}]{zhou2020simulations}
Zhou, Y., Chen, P., Hong, J., \& Fang, C. 2020, Nature Astronomy, 4, 994

\end{thebibliography}

\listofchanges

\end{CJK*}

\end{document}